\documentclass[a4paper,12pt]{article}
\usepackage[utf8]{inputenc}
\usepackage[hidelinks]{hyperref}
\usepackage[margin=2cm]{geometry}
\usepackage{amsmath,amssymb}
\usepackage{authblk}
\usepackage{bm}
\usepackage{graphicx}
\usepackage[title]{appendix}
\usepackage{lineno}
\usepackage{listings}
\usepackage{color}

\newcommand{\rc}[1]{{\color{green} #1}}
\usepackage[normalem]{ulem}

\newcommand{\matlab}{Matlab\textsuperscript{\textregistered}}

\title{A nearest-neighbour discretisation of the regularized stokeslet boundary
integral equation}
\author{David J. Smith\thanks{D.J.Smith@bham.ac.uk}}
\affil{School of Mathematics, University of Birmingham, Edgbaston, Birmingham,
B15 2TT, UK}
\date{}

\begin{document}

\maketitle

\lstset{language=matlab}

\begin{abstract}
The method of regularized stokeslets is extensively used in biological fluid
dynamics due to its conceptual simplicity and meshlessness. This simplicity carries a degree of cost in computational expense and accuracy because the number of degrees of freedom used to
discretise the unknown surface traction is generally significantly higher than that required by boundary element methods. We describe a
meshless method based on nearest-neighbour interpolation that significantly reduces
the number of degrees of freedom required to discretise the unknown traction, increasing the range of problems that can be practically solved, without excessively complicating the task of the modeller.
The nearest-neighbour technique is tested against the classical problem of rigid body motion of a sphere immersed in very viscous fluid, then applied to the more complex biophysical problem of
calculating the rotational diffusion timescales of a macromolecular structure modelled by three
closely-spaced non-slender rods.
A heuristic for finding the required density of force and quadrature points by
numerical refinement is suggested. Matlab/GNU Octave code for the key steps of the algorithm is provided,
which predominantly use basic linear algebra operations, with a full implementation
being provided on github. Compared with the standard Nystr\"{o}m discretisation, more accurate and substantially more efficient results can be obtained by
de-refining the force discretisation relative to the quadrature discretisation: a cost reduction of
over 10 times with improved accuracy is observed. This improvement comes at minimal additional technical complexity. Future avenues to develop the algorithm are then discussed.
\end{abstract}

\section{Introduction}
When attempting to formulate and solve mathematical models of microscopic
biological flow systems, for example involving macromolecular structures,
swimming cells and cilia, a significant challenge to overcome is that the flow domain is
typically bounded by curved, moving surfaces. Often it is of interest to model line-like
objects such as cilia and flagella, and point-like bodies such as suspensions of
many bacteria, in addition to genuinely 2D surfaces. The Stokes flow equations
are linear, and in some celebrated cases it has been found possible to make
significant analytical progress, for example by exploiting small amplitude
expansions in the boundary movement \cite{taylor1951} or slenderness
\cite{hancock1953,gray1955,chwang1971}, for certain idealised problems (for a
more detailed review of the field, see Lauga \& Powers \cite{lauga2009}). However
the majority of problems of practical interest, typically involving multiple cells, non-planar domains and large
amplitude motions, require computational modelling, and there has been intensive activity
in this area in the last decade.

The linearity of the flow equations enables the formulation of methods based
on the boundary integral equation for Stokes flow; these methods remove the need
to discretise and solve directly in the flow volume, as would be necessary for
the finite element method. This reduction in dimensionality both removes the
need to mesh and re-mesh the evolving flow domain, and vastly reduces the size
of the linear algebra problem resulting from discretisation. In certain respects
these methods were anticipated by the computational slender body theory work of
Higdon \cite{higdon1979} and \cite{johnson1979}; relatively early examples of
the `fully-fledged' boundary element method for Stokes flow was developed by
Phan-Thien and colleagues \cite{phan1987,ramia1993}. The achievements of the latter
group with late 1980s/early 1990s computational hardware set a benchmark for
work in the current era of desktop machines with multi-gigabyte RAM. It should
  of course be noted that there have been major algorithmic developments in numerical methods
  for Stokes flow in the intervening period, including the
  completed double-layer boundary
  integral equation \cite{power1987,klinteberg2016}, hybrid boundary integral-multipole
  methods \cite{zinchenko2000}, spectral discretisation combined with the fast multipole
  method \cite{veerapaneni2009,veerapaneni2011}, quadrature by expansion \cite{klinteberg2016},
  and slender body theory combined with these techniques \cite{nazockdast2017}. These approaches
  are generally employed by numerical experts to solve problems at the limits of
  computational feasibility, involving very large numbers of interacting bodies.

  The classical boundary element method for Stokes flow, along with the more advanced methods
    described above, are both accurate and
  efficient. However, they present two technical challenges in their implementation, particularly when considered from
    the point of view of users who are not computational specialists. The first challenge is
the need to generate a surface \emph{mesh}, i.e.\ a geometric discretisation of
all surfaces in the problem consisting of oriented smooth, and
smoothly-connected, patches which interpolate several surface points.\footnote{In this paper the term
    \emph{mesh} will be reserved for an object \((P,E)\) where \(P=\{\bm{x}[1],\ldots,\bm{x}[N]\}\in\mathbb{R}^3\) is an ordered set of \emph{points/nodes}, and \(E\) is a table defining the \emph{elements} of the mesh, e.g.\ for a mesh of flat triangles, the elements take the form \((\bm{x}[E(1,e)],\bm{x}[E(2,e)],\bm{x}[E(3,e)])\). Where we refer to a set \(P\) without the associated table defining the elements, the terms \emph{discretisation} or \emph{points} will be used instead. The aim of this study is to achieve improved accuracy and efficiency without needing to construct \(E\).}  While much
easier than the volumetric meshing that would be required for the finite element
method, meshes of even moderately complicated biomolecular or cellular structures may require
significant time and ingenuity to create, and may not be suited to automated
generation -- as might be needed to study biological heterogeneity. Furthermore,
some objects will appear to a very good approximation as lines or points --
detailed surface meshing of these bodies may involve a level of computational
refinement that is unwarranted. The second challenge -- which has arguably been
addressed through the availability of library code such as BEMLIB
\cite{pozrikidis2002} -- is the singularity of the stokeslet velocity and stress
kernels, and requirement for semi-analytical quadrature methods. The latter
issue does however present an additional layer of complexity for those who are
not numerical specialists.

The method of regularized stokeslets, introduced by Cortez and colleagues
\cite{cortez2001,cortez2005,ainley2008,cortez2014}, has proved to be an
effective and accessible method for simulating and analysing microscale biological flows. This
method deals effectively with both of the above difficulties by removing the
need for a true mesh, requiring only a set of discrete points approximating the solid
objects in the flow, and regularising the integral kernel so that specialised
quadrature is not required. The core idea is the derivation of a family of
regularized versions of the singular stokeslet/Oseen tensor kernel that
nevertheless satisfy exact conservation of mass. Whereas the singular stokeslet
corresponds to the Stokes flow produced by a Dirac delta force-per-unit-volume
distribution, a regularized stokeslet corresponds to the Stokes flow produced by
a `blob', i.e.\ a finite force-per-unit-volume distribution which approximates a
Dirac delta function. Cortez and colleagues have derived various versions of the
regularized stokeslet corresponding to both 2D \cite{cortez2001} and 3D
\cite{cortez2005} domains, to various forms of blob distribution
\cite{cortez2001}, with image systems to represent a plane boundary
\cite{ainley2008,cortez2015}, and for periodic problems \cite{cortez2014}. We
will not attempt to give a comprehensive survey of applications of the method of
regularized stokeslets; it suffices to note that a Google Scholar search on 28th
April 2017 with the term ``regularized stokeslets'' produced 250 results since
2012.

The standard numerical implementation of the method of regularized stokeslets is
to employ a Nystr\"{o}m discretisation of the Fredholm integral equation, which
replaces the integral directly with a quadrature rule. This method is very
simple to implement, and has been used in the great majority of published work.
This simplicity does however come at a computational cost, arising from the
fact that the quantity of interest in a boundary integral equation method, the
surface traction distribution, varies much more slowly than the near-singular
kernel. Therefore very many degrees of freedom, corresponding to the
discretisation of the traction, are required in order for the quadrature to be
accurate. Furthermore, there is a coupling between the discretisation length
scale and the regularisation parameter that must be satisfied in order for
results to be considered converged. As a consequence, the RAM requirements alone
for relatively simple geometries may be very high, as evident in a number of
recent studies on helical flagella for example.

The issue of the computational cost of the method of regularized stokeslets was
discussed in an earlier paper \cite{smith2009}, in which we suggested employing a
boundary element discretisation of the regularized stokeslet boundary integral
equation. This approach is undoubtedly computationally efficient, and formed the
basis for subsequent detailed modelling of the left right organising structures
of mouse \cite{smith2011} and zebrafish \cite{smith2012,sampaio2014}, however it
becomes necessary to generate a mesh in the same way as the classical boundary
element method.

In this paper we will describe an alternative `nearest-neighbour' discretisation
of the method of regularized stokeslets which retains the meshless simplicity of
the standard approach, but has greatly reduced computational cost. Alongside the
mathematical description, an implementation in \matlab/GNU Octave will be given,
and applied to a simple test problem of the drag and moment on a sphere or prolate spheroid undergoing
rigid body motion, followed by a more complex problem of calculating the rotational diffusion
timescale of a biological macromolecule.

\section{Stokeslets and boundary integral methods}
The very low Reynolds numbers associated with microscopic flows on the length
scales of macromolecules and cells motivates the study of the Stokes flow equations for
viscous-dominated flow. The dimensionless form of these equations is,
\begin{equation}
-\bm{\nabla}p+\nabla^2\bm{u}=0, \quad \nabla \cdot \bm{u}=0,\label{eq:stokes}
\end{equation}
augmented with the no-slip, no-penetration boundary condition
\(\bm{u}(\bm{X})=\dot{\bm{X}}\) for boundary points \(\bm{X}\). The basis
for boundary integral and singularity methods is to exploit the linearity of
eq.~\eqref{eq:stokes} to construct solutions satisfying the required boundary
conditions from sums and/or integrals of fundamental solutions.

The classical singular fundamental solution is the stokeslet or Oseen tensor,
given by the second rank tensor \(S_{jk}\) and first rank tensor \(P_k\) for
which \(\bm{u}=(S_{1k},S_{2k},S_{3k})\) and \(p=P_k\) are the solutions of the
Stokes flow equations with a Dirac delta distribution force-per-unit-volume located at \(\bm{y}\):
\begin{equation}
-\bm{\nabla}p+\nabla^2\bm{u}+8\pi\bm{e}_k\delta(\bm{x}-\bm{y})=0, \quad \nabla \cdot
\bm{u}=0.
\end{equation}
The form of the stokeslet in 3D is,
\begin{align}
S_{jk}(\bm{x},\bm{y})& =
\frac{\delta_{jk}}{|\bm{x}-\bm{y}|}+\frac{(x_j-y_j)(x_k-y_k)}{|\bm{x}-\bm{y}|^3}\rc{,}
 \\
P_k(\bm{x},\bm{y})   & = 2\frac{x_k-y_k}{|\bm{x}-\bm{y}|^3}.
\end{align}

The singularity method for Stokes flow involves seeking an approximate solution
to equation~\eqref{eq:stokes} by locating Stokeslets, and sometimes higher
order stokes-multipoles, outside of the flow domain. For example, singularities may be located inside cells, or along the centrelines of cilia and flagella as in slender body theory; the simplest example is perhaps the solution to Stokes flow driven by a translating sphere, which can be expressed as the sum of a stokeslet and source-dipole (the latter being a special case of the stokes-quadrupole) at the centre of the sphere. Review and references are given for example Smith et al.\ \cite{smith2009}.

Conversely, the boundary integral method for Stokes flow involves formulating the exact integral equation,
\begin{equation}
 u_j(\bm{y}) = -\frac{1}{8\pi} \iint_{\partial D}
  S_{ij}(\bm{x},\bm{y})f_i(\bm{x}) dS(\bm{x}) + \frac{1}{8\pi}\iint_{\partial D} u_i(\bm{x})T_{ijk}(\bm{x},\bm{y})n_k(\bm{x}) dS(\bm{x}), \label{eq:bie}
\end{equation}
where \(T_{ijk}\) is the stress tensor associated with the Stokes flow \(\bm{u}=(S_{1k},S_{2k},S_{3k})\), \(p=P_k\), given by
\begin{equation}
T_{ijk}(\bm{x},\bm{y})=-\frac{6(x_i-y_i)(x_j-y_j)(x_k-y_k)}{|\bm{x}-\bm{y}|^5}.
\end{equation}
The summation convention for repeated indices is used throughout.
The boundary integral equation is solved numerically by taking the limit of equation~\eqref{eq:bie} as \(\bm{y}\) approaches the bounding surfaces of the domain from within the fluid, then performing discretisation of the surface geometry \(\partial D\) and traction \(\bm{f}\). If the boundary of the domain is stationary and immersed objects in the domain are rigid bodies, the `double layer' term arising from the integral of the stress is identically zero and so the flow is given exactly by a surface distribution of stokeslets only; under the weaker condition that \(\iint_{\partial D} \bm{u}\cdot\bm{n}\, dS=0\) it can also be shown that the double layer integral may be eliminated by taking a modified Stokeslet density, which is no longer precisely the surface traction. In either case, the flow is given exactly by boundary integrals of `single layer' stokeslet velocity tensors only \cite{pozrikidis1992}.

A detailed exposition of the boundary element method for Stokes flow and its numerical implementation is given by Pozrikidis \cite{pozrikidis1992,pozrikidis2002}. The boundary integral and singularity methods may be hybridised to formulate approximate but accurate and efficient simulation of cell movement \cite{smith2009sperm}.

The integral equation problem formed from equation \eqref{eq:bie} in the limit \(\bm{y}\rightarrow \bm{Y}\in\partial D\) possesses singular integrals which require specialised evaluation; moreover line and point singularity distributions, while they may not lie strictly in the flow domain, may nevertheless complicate the evaluation of flow fields for purposes such as particle tracking. An additional complication for boundary element methods is the requirement to build a true surface mesh.  It should be emphasised that these issues are technical complications rather than inherent problems, however methods which do not possess these complications are appealing, particularly for biological flow, as evidenced by the rapid adoption and use of the method of regularized stokeslets, which we will briefly review in the next section.

\section{The method of regularized stokeslets and its numerical implementation}
Cortez \cite{cortez2001} formulated the regularized stokeslet as the exact solution to the incompressible Stokes flow equations forced by a spatially-smoothed force per unit volume, \(\phi_\epsilon(\bm{x}-\bm{y})\),
\begin{equation}
-\bm{\nabla}p+\nabla^2\bm{u}+8\pi\bm{e}_k\phi_\epsilon(\bm{x}-\bm{y})=0, \quad \nabla \cdot
\bm{u}=0.
\end{equation}
The `blob' \(\phi_\epsilon\) denotes a family of functions parameterised by \(\epsilon\) satisfying \(\int \dots \int_{\bm{R}^n} \phi_\epsilon dV = 1\), and tending to a Dirac delta distribution in the limit \(\epsilon\rightarrow 0\). The derivation of specific forms of the regularized stokeslet were discussed by Cortez and colleagues \cite{cortez2001,cortez2005}; we will suffice by noting that a frequently-used form for 3D flow is based on the blob function,
\begin{equation}
 \phi_\epsilon(\bm{\xi})=\frac{15\epsilon^4}{8\pi(|\bm{\xi}|^2+\epsilon^2)^{7/2}}, \label{eq:blob}
\end{equation}
which leads to the regularized Stokeslet pressure and velocity tensors,
\begin{align}
  P_j^\epsilon(\bm{x},\bm{y})     &= (x_j-y_j)\frac{2|\bm{x}-\bm{y}|^2+5\epsilon^2}{(|\bm{x}-\bm{y}|^{2}+\epsilon^2)^{5/2}}, \\
  S_{ij}^\epsilon(\bm{x},\bm{y})  &= \delta_{ij}\frac{|\bm{x}-\bm{y}|^2+\epsilon^2}{(|\bm{x}-\bm{y}|^{2}+\epsilon^2)^{3/2}}+\frac{(x_i-y_i)(x_j-y_j)}{(|\bm{x}-\bm{y}|^{2}+\epsilon^2)^{3/2}} , \\
  T_{ijk}^\epsilon(\bm{x},\bm{y})  &= -\frac{6(x_i-y_i)(x_j-y_j)(x_k-y_k)}{(|\bm{x}-\bm{y}|^2+\epsilon^2)^{5/2}}
                                      -\frac{3\epsilon^2[(x_i-y_i)\delta_{jk}+(x_j-y_j)\delta_{ik}+(x_k-y_k)\delta_{ij}]}{(|\bm{x}-\bm{y}|^2+\epsilon^2)^{5/2}}.
\end{align}

The regularized counterpart to the classical boundary integral equation~\eqref{eq:bie} in 3D is,
\begin{align}
 u_j(\bm{y})&\approx \iiint_{\mathbb{R}^3} u_j(\bm{x}) \phi_\epsilon(\bm{x}-\bm{y}) dV(\bm{x}) \nonumber \\
 & = -\frac{1}{8\pi} \iint_{\partial D} S_{ij}^{\epsilon}(\bm{x},\bm{y})f_i(\bm{x}) dS(\bm{x})
   - \frac{1}{8\pi} \iint_{\partial D} u_i(\bm{x})T_{ijk}^{\epsilon}(\bm{x},\bm{y})n_k(\bm{x}) dS(\bm{x}).\label{eq:rbie}
\end{align}
Unlike the classical boundary integral equation, the regularized version~\eqref{eq:rbie} is approximate even before the numerical discretisation is carried out; for the blob function~\eqref{eq:blob} the error is \(O(\epsilon^2)\) for \(\bm{y}\) greater than distance \(\sqrt{5\epsilon/2}\) from the boundary, and \(O(\epsilon)\) otherwise \cite{cortez2005}. The double layer integral is typically eliminated in practical implementations of the regularized stokeslet. This elimination may be formally justified for boundaries undergoing rigid body motion, for example models of spirochetes as rotating helices \cite{cortez2005} and cilia undergoing purely rotational motion \cite{smith2012}, however for bodies which undergo significant flexible motion such as respiratory cilia and sperm flagella, this elimination is an approximation which must be justified by either post hoc numerical checks \cite{smith2009} or slender body theory analysis \cite{cortez2012}. The resulting approximate single-layer boundary integral equation is then,
\begin{align}
 u_j(\bm{y}) & \approx -\frac{1}{8\pi} \iint_{\partial D} S_{ij}^{\epsilon}(\bm{x},\bm{y})f_i(\bm{x}) dS(\bm{x}).\label{eq:rslbie1}
\end{align}
In what follows we will treat the approximation as exact, however it should be borne in mind that there is error associated with both the continuous integral equation~\eqref{eq:rslbie1} in addition to the error associated with subsequent discretisation. In what follows we will find it convenient to use the identity \(S_{ij}^\epsilon(\bm{x},\bm{y}) = S_{ji}^\epsilon(\bm{y},\bm{x})\); relabelling, and treating the approximation as exact we have,
\begin{align}
 u_i(\bm{x}) & = -\frac{1}{8\pi} \iint_{\partial D} S_{ij}^{\epsilon}(\bm{x},\bm{y})f_j(\bm{y}) dS(\bm{y}).\label{eq:rslbie}
\end{align}

If the body motion is prescribed, the no-slip condition \(\bm{u}(\bm{x})=\dot{\bm{x}}\) can be applied on the surface \(\partial D\) to convert equation~\eqref{eq:rslbie} to a Fredholm first kind integral equation for the unknown force distribution \(\bm{f}(\bm{y})\) -- a \emph{resistance problem}.
\begin{align}
 \dot{x}_i& = -\frac{1}{8\pi} \iint_{\partial D} S_{ij}^{\epsilon}(\bm{x},\bm{y})f_j(\bm{y}) dS(\bm{x}) \quad \mbox{all} \quad \bm{x}\in\partial D.\label{eq:resprob}
\end{align}

If the body is rigid, or its surface velocity is known up to a rigid body motion, and the total force and moment \(\bm{F}\), \(\bm{M}\) are known, the result is the \emph{mobility problem},
\begin{align}
  \dot{x}_i+U_i+\epsilon_{ijk}\Omega_j x_k & = -\frac{1}{8\pi} \iint_{\partial D} S_{ij}^{\epsilon}(\bm{x},\bm{y})f_j(\bm{y}) dS(\bm{y}) \quad \mbox{all} \quad \bm{x}\in\partial D,\nonumber \\
  F_i & = \iint_{\partial D} f_i (\bm{y})dS(\bm{y}), \nonumber \\
  M_i & = \iint_{\partial D}\epsilon_{ijk}y_j f_k (\bm{y}) dS(\bm{y}),\label{eq:mobprob}
\end{align}
where the rigid body velocity \(\bm{U}\) and angular velocity \(\bm{\Omega}\), and the force distribution \(\bm{f}(\bm{y})\), are unknown; \(\epsilon_{ijk}\) is the Levi-Civita alterating tensor. The mobility problem arises from situations such as a sedimenting body (for which the force is given by gravity or centrifugal force and the moment is zero), or a swimming cell in the inertialess regime of Stokes flow (for which the force and moment are both zero).

To solve the problems~\eqref{eq:resprob} and \eqref{eq:mobprob}, the method of numerical discretisation described by Cortez et al.\ \cite{cortez2005} and used in the majority of studies to date takes advantage of the regularity of the \(S_{ij}^\epsilon\) kernel and directly approximates the surface integrals with a quadrature rule followed by collocation on the quadrature points. The result is a system such as,
\begin{equation}
  \dot{x}_i[m] = \frac{1}{8\pi}\sum_{n=1}^N S_{ij}^\epsilon(\bm{x}[m],\bm{x}[n])g_j[n] A[n],
\end{equation}
for the resistance problem, where \((\bm{x}[n],A[n])\) are quadrature nodes and weights, and \(g_j[n]=-f_j(\bm{x}[n])\). For the mobility problem, we have,
\begin{align}
  \dot{x}_i[m] & = \frac{1}{8\pi}\sum_{n=1}^N S_{ij}^\epsilon(\bm{x}[m],\bm{x}[n])g_j[n] A[n],  \quad \mbox{for} \quad m=1,\ldots, N,\nonumber \\
  F_i & = \sum_{n=1}^N g_i[n] A[n], \nonumber \\
  M_i & = \sum_{n=1}^N \epsilon_{ijk}x_j[n]g_k[n] A[n].
\end{align}
The above approach has the principal advantage of computational simplicity, and the principal disadvantage that the degrees of freedom of the resulting linear system are tied to the quadrature required to approximate the rapidly-varying kernel \(S_{ij}^\epsilon(\bm{x},\bm{X})\) for \(|\bm{x}-\bm{X}|=O(\epsilon)\) -- and associated high computational expense for a given level of accuracy.

Boundary element methods take an alternative approach to numerical discretisation -- to discretise the unknown density \(\bm{f}(\bm{y})\) with basis functions \(\Phi_n(\bm{y})\), i.e.\ \(\bm{f}(\bm{y})=-\sum_{n=1}^N \bm{g}[n]\Phi_n(\bm{y})\). The integral operator can then be written as,
\begin{equation}
 - \iint_{\partial D} S_{ij}^{\epsilon}(\bm{x},\bm{y})f_j(\bm{y}) dS(\bm{y}) = \sum_{n=1}^N g_j[n] \iint_{\partial D} S_{ij}^{\epsilon}(\bm{x},\bm{y}) \Phi_n(\bm{y}) dS(\bm{y}).
\end{equation}
In the simplest `constant force' implementation, the basis functions \(\{\Phi_1,\ldots,\Phi_N\}\) are indicator functions on the elements of the mesh \(\{E_1,\dots,E_N\}\).
The stokeslet integrals are then decoupled from the force discretisation, and can be subjected to suitably fine spatial discretisation as appropriate, without unnecessarily increasing the number of degrees of freedom in the system -- a major saving in both computational storage and time. This approach was suggested in the context of regularized stokeslet methods by Smith \cite{smith2009}, and subsequently applied to problems in developmental biology \cite{smith2012,sampaio2014} and sperm cell motion \cite{montenegro2015}. The practical drawback of this method is the need to generate a true surface mesh, which for complex geometries may be time-consuming.

To retain the advantages of both approaches -- ease of implementation and computational efficiency -- we suggest an alternative approach based on nearest-neighbour interpolation.

\section{Nearest-neighbour discretisation of the regularized stokeslet boundary integral}
Suppose that we have two surface discretisations of \(\partial D\), \(\{\bm{x}[1],\ldots,\bm{x}[N]\}\) and \(\{\bm{X}[1],\ldots,\bm{X}[Q]\}\), which we will refer to as the \emph{force discretisation} and \emph{quadrature  discretisation} respectively. These discretisations are not true meshes because they are not equipped with a mapping from nodes to elements, and we will not need to evaluate integrals in local coordinate systems. In general, \(N\ll Q\) because the kernel \(S_{ij}^\epsilon(\bm{x},\bm{y})\) varies much more rapidly than the surface traction \(\bm{f}(\bm{y})\).

Provided that they do not vary rapidly relative to the force points, the force \(\bm{f}(\bm{y})\) and surface metric \(dS(\bm{y})\) may then be discretised using nearest-neighbour interpolation. Denote by \(\mathcal{N}:\{1,\ldots,Q\}\rightarrow \{1,\ldots,N\}\) the nearest-neighbour discretisation such that,
\begin{equation}
  \mathcal{N}(q):=\underset{n=1,\ldots,N}{\mbox{argmin}} \, |\bm{x}[n]-\bm{X}[q]|,
\end{equation}
so that \(f_j(\bm{X}[q]) dS(\bm{X}[q]) \approx f_j(\bm{x}[\mathcal{N}(q)]) dS(\bm{x}[\mathcal{N}(q)])=:-g_j[\mathcal{N}(q)] A[\mathcal{N}(q)]\). The nearest-neighbour operator \(\mathcal{N}\) can be expressed as a \(Q\times N\) matrix,
\begin{equation}
  \nu[q,\hat{n}]=\begin{cases} 1 \quad \mbox{if} \quad \hat{n} = \underset{n=1,\ldots,N}{\mbox{argmin}} \, |\bm{x}[n]-\bm{X}[q]|, \\ 0 \quad \mbox{otherwise}, \end{cases}
\end{equation}
so that \(g_i[\mathcal{N}(q)] A[\mathcal{N}(q)]=\sum_{n=1}^N \mathsf{\nu}[q,n] g_i[n] A[n]\).

With the above discretisation, the regularized stokeslet boundary integral may be approximated as,
\begin{align}
  -\iint_{\partial D} S_{ij}^{\epsilon}(\bm{x},\bm{y})f_j(\bm{y}) dS(\bm{y}) & \approx -\sum_{q=1}^Q S_{ij}^{\epsilon}(\bm{x},\bm{X}[q]) f_j(\bm{x}[\mathcal{N}(q)]) A[\mathcal{N}(q)], \nonumber \\
  & = \sum_{q=1}^Q S_{ij}^{\epsilon}(\bm{x},\bm{X}[q]) \sum_{n=1}^N \mathsf{\nu}[q,n] g_j[n] A[n].  \label{eq:nndisc}
\end{align}
Applying the discretisation~\eqref{eq:nndisc} to the boundary integral equation~\eqref{eq:rslbie}, followed by performing collocation on the force discretisation \(\bm{u}(\bm{x}[m])=\dot{\bm{x}}[m]\), leads to the discretised resistance problem,
\begin{equation}
  \dot{x}_i[m] = \frac{1}{8\pi} \sum_{n=1}^N g_j[n] A[n] \sum_{q=1}^Q S_{ij}^\epsilon(\bm{x}[m],\bm{X}[q]) \mathsf{\nu}[q,n] , \label{eq:nnres}
\end{equation}
and mobility problem,
\begin{align}
  \dot{x}_i[m] & =  \frac{1}{8\pi} \sum_{n=1}^N g_j[n] A[n] \sum_{q=1}^Q S_{ij}^\epsilon(\bm{x}[m],\bm{X}[q]) \mathsf{\nu}[q,n],  \quad \mbox{for} \quad m=1,\ldots, N, \nonumber \\
  F_i       & = \sum_{n=1}^N g_i[n]  A[n] \sum_{q=1}^Q \nu[q,n] , \nonumber \\
  M_i       & = \sum_{n=1}^N g_k[n]  A[n] \sum_{q=1}^Q \epsilon_{ijk}X_j[q] \nu[q,n] . \label{eq:nnmob}
\end{align}

The discrete resistance problem~\eqref{eq:nndisc} can be written as a \(3N\times 3N\) linear system \(\mathsf{A}\mathsf{f}=\mathsf{b}\), where the unknown \(3N\)-vector \(\mathsf{f}\) has components,
\begin{equation}
\mathsf{f}[N(j-1)+n]=g_j[n] A[n] , \label{eq:nnunknown}
\end{equation}
the \(3N\times 3N\) left hand side matrix \(\mathsf{A}\) has components,
\begin{equation}
\mathsf{A}[N(i-1)+m,N(j-1)+n]=\frac{1}{8\pi} \sum_{q=1}^Q S_{ij}^\epsilon(\bm{x}[m],\bm{X}[q]) \sum_{n=1}^N \mathsf{\nu}[q,n], \label{eq:nnres2}
\end{equation}
and the right hand side velocity is given by,
\begin{equation}
\mathsf{b}[N(i-1)+m]=\dot{x}_i(\bm{x}[m]). \label{eq:nnrhs}
\end{equation}

The discrete mobility problem can be written similarly as a \(3(N+2)\times 3(N+2)\) linear system, where \(\mathsf{A}\) and \(\mathsf{b}\) are augmented by six rows discretising the force and moment constraints, and \(\mathsf{f}\) has six additional scalar unknowns representing the values of \(\bm{U}\) and \(\bm{\Omega}\).

The discrete problems~\eqref{eq:nnres} and \eqref{eq:nnmob} may be implemented in \matlab or GNU Octave by assembling matrices representing \(S_{ij}^\epsilon(\bm{x}[m],\bm{X}[q])\) and \(\nu[q,n]\). Details are provided in appendix~\ref{app:code}.

\section{Numerical results and analysis}
The core numerical codes for implementation of the method given by equations~\eqref{eq:nnunknown}--\eqref{eq:nnrhs} are given in appendices~\ref{app:regsto}--\ref{app:resprob}. The full code (approximately 1000 lines) used to produce the results in this report is available from github at \url{https://github.com/djsmithbham/NearestStokeslets}. The quadrature weights are absorbed into the \(g_i[n]\) and so are never calculated explicitly.

For numerical testing we will denote the maximum discretisation spacing (i.e.\ maximum distance of a point to its nearest-neighbour) by \(h_f\) for the force points and \(h_q\) for the quadrature points:
\begin{align}
  h_f & = \max_{m=1,\ldots,N} \min_{\substack{n=1,\ldots,N\\n\not = m}} |\bm{x}[m]-\bm{x}[n]|\nonumber \\
  h_q & = \max_{p=1,\ldots,Q} \min_{\substack{q=1,\ldots,Q\\q\not = p}} |\bm{x}[p]-\bm{x}[q]|.
\end{align}
This parameter may be computed for a given discretisation as described in appendix~\ref{app:calcdiscr_h}.

\subsection{Rigid body motion of a sphere}
The simplest test case is perhaps Stokes' law for a translating or rotating sphere in an infinite fluid. Taking a sphere of radius \(1\) translating with velocity \(\bm{U}=(1,0,0)\), the exact solution to the resistance problem yields total force \(\bm{F}=(6\pi,0,0)\); rotation with velocity \(\bm{\Omega}=(1,0,0)\) yields total moment \(\bm{M}=(8\pi,0,0)\). Discretising the sphere by projecting onto the six faces of a cube yields the discretisations shown in figure~\ref{fig:sphereDiscr} (a -- force/collocation points, b -- quadrature points).

Numerical experiments assessing the \(L_2\) relative error in total force and moment compared with analytic solutions, for varying regularisation parameter \(\epsilon\), force points \(h_f\) and quadrature points \(h_q\), are shown in tables~ \ref{tab:sph001}, \ref{tab:sph002} and \ref{tab:sph0005} and example computational timings are given in appendix~\ref{app:sphTimings}, table~\ref{tab:sphTimings}. The entries on the main diagonal (\(h_f=h_q\)) correspond to the Nystr\"{o}m discretisation; `non-trivial' nearest-neighbour results are above the main diagonal (\(h_f<h_q\)). Results below the main diagonal correspond to more force points than quadrature points; in all cases the system is ill-conditioned (table~\ref{tab:sphCond}) and the \matlab linear solver returns `NaN' (not-a-number). Conditioning is generally not a problem provided that \(h_f< h_q\), or if \(h_f=h_q\) and the force and quadrature discretisations coincide. If \(h_f=h_q\) and the discretisations are non-overlappling, singular matrices can result -- data not shown).

It is immediately clear from examining the table rows that for fixed force discretisation spacing \(h_f\), decreasing the quadrature discretisation spacing \(h_q\) typically results in improved accuracy, notwithstanding a slight reversal in this tendency which may occur for very coarse \(h_f=0.58\) and very fine \(h_q<0.02\). This behaviour can be interpreted as progressively finer \(h_q\) enabling more progressively more accurate quadrature, until the error is instead dominated by errors associated with force discretisation. An error estimate will follow in section~\ref{sec:errorEstimate}.

Examining the columns of tables~\ref{tab:sph001} (see also tables~\ref{tab:sph002} and \ref{tab:sph0005}) reveals a more interesting behaviour of the algorithm. If the quadrature discretisation size \(h_q\) is fixed, more accurate results are obtained with the force spacing \(h_f\) taken \emph{coarser} than the quadrature spacing (\(h_f>h_q\)) than with the Nystr\"om method (\(h_f=h_q\)). Appendix \ref{app:sphTimings} confirms that, for fixed \(h_q\), the choice \(h_f=h_q\) can be rather inaccurate, and is sensitive to the value of \(\epsilon\), whereas taking \(h_f\approx 2h_q\) reliably produces results which are accurate to within a few percent, and at much lower computational cost (see appendix~\ref{app:sphTimings}). A similar result is observed for the slightly more complex problem of calculating the resistance tensor of a prolate spheroid (appendix \ref{app:prolate}).

The effect of the regularisation parameter is discussed in appendix ~\ref{app:calcdiscr_h}. Reducing \(\epsilon\) typically reduces the error for all finite \(\epsilon\) tested, provided that \(h_q<h_f/2\). The regularisation error is proportional to \(\epsilon\), however it may be expected that as \(\epsilon\) is reduced, \(h_q\) may have to be reduced proportionately in order to approximate the integral of the increasingly-peaked kernel more accurately. However, this behaviour was not observed in the test cases analysed (for which \(\epsilon\) was taken as small as \(10^{-6}\)). In applications in which evaluation of the velocity field is of interest, a balance between small regularisation error and smooth/efficient evaluation of the velocity field may be sought, motivating an intermediate choice of \(\epsilon\).

The final quantity to consider is the force discretisation length \(h_f\). This discretisation must be fine enough to resolve variations in the surface force density. The translating sphere case in fact is not a good way to assess this convergence, because the surface stress is constant \cite[p.\ 233]{batchelor1967}! The rotating sphere does however possess a non-constant surface force density, which varies from zero at the poles to its maximum at the equator. From the results in tables~\ref{tab:sph002}--\ref{tab:sph0005} it is clear that the coarsest force discretisation \(h_f=0.58\) produces acceptably accurate results (i.e.\ within about 1\% error) provided that the quadrature discretisation is sufficiently fine.

\subsection{Error estimate}\label{sec:errorEstimate}
Following these numerical experiments, we shall briefly outline an error estimate for the nearest-neighbour method. There are three sources of error: (i) regularisation error associated with the use of the regularised version of the boundary integral equation with parameter \(\epsilon\) -- which was discussed above following equation~\eqref{eq:rbie}, (ii) discretisation error associated with the approximation of the integral by its values on the quadrature points, which have spacing \(h_q\), (iii) discretisation error associated with the approximation of the force and metric by their values on the coarser force points, which have spacing \(h_f\).

The discretisation error associated with the approximation of the integral by its values on the quadrature points will be chiefly determined by the contribution associated with the rapid variation in the kernel. We will restrict to the case where \(h_f\gg \epsilon\). The lowest order estimate of quadrature error follows from taking the mean value inequality, i.e.\ \(|S_{jk}(\bm{x},\bm{y})-S_{jk}(\bm{x},\bm{X}[q])|\leqslant M_1|(\bm{y}-\bm{X}[q])|\), where \(M_1\) is a bound on \(|\bm{\nabla}_{\bm{y}} S_{jk}(\bm{x},\bm{X}[q])|\). The integrand is sharply-peaked but in a small area -- to take account of this behaviour more precisely, the integral will be split into three regions based on the value of \(r=|\bm{x}-\bm{y}|\), the regions (i) \(0<r<h_f\), (ii) \(h_f<r<h_f^{1/2}\) and (iii) \(h_f^{1/2}<r\), and the error estimated on each region in turn and summed.

\sloppy
\begin{enumerate}
\item [(i)]
  Considering first the `near' part of the integral encountered around the collocation point, i.e.\ where \(|\bm{x}-\bm{y}|\leqslant h_f\), and noting that the regularised stokeslet is dominated by the behaviour of \((r^2+\epsilon^2)^{-1/2}\), the bound \(M_1=O(\epsilon^{-2})\) and so the error in the surface integral is \(O(\epsilon^{-2} h_f^2 h_q)\), because the area of the region is \(O(h_f^2)\) and the spacing between collocation points is \(O(h_q)\).

\item [(ii)]
  In the intermediate region \(h_f<|\bm{x}-\bm{y}|\leqslant h_f^{1/2}\), the bound \(M_1=O(h_f^{-2})\) over an area \(O(h_f)\), yielding a quadrature error \(O(h_f^{-1} h_q)\).
\item [(iii)]
  For the outer region \(h_f^{1/2}<|\bm{x}-\bm{y}|\), the bound \(M_1=O(h_f^{-1})\) and the area is \(O(1)\), giving a quadrature error \(O(h_f^{-1}h_q)\) again.
\end{enumerate}
The total discretisation error associated with quadrature can therefore be estimated as \(O(\epsilon^{-2} h_f^2 h_q) + O(h_f^{-1} h_q)\). The first term may not be a sharp estimate; the results of table~\ref{tab:sph1e-6} suggest that accurate results may be obtained (perhaps for certain types of discretisation) for very small \(\epsilon\) compared with \(h_f\) and \(h_q\). The second term emphasises the advantage of taking \(h_f>h_q\), i.e.\ the force points coarser than the quadrature points.

Finally, the discretisation error associated with the approximation of the force and metric by their values on the force points can be estimated by noting that the error of nearest-neighbour interpolation is again of the form \(M_2|\bm{X}[q]-\bm{x}[\mathcal{N}(q)]|\), where \(M_2\) is a bound on \(\|\nabla_{\bm{y}} (\bm{f}(\bm{y}) dS(\bm{y}))\|\). Hence the force discretisation error is \(O(h_f)\).

In summary, our estimate of the error associated with the regularisation and nearest-neighbour discretisation of the boundary integral equation is \(O(\epsilon)+O(\epsilon^{-2} h_f^2 h_q) + O(h_f^{-1} h_q)+O(h_f)\). The numerical results are consistent with the finding that there are independent errors due to regularisation (see appendix~\ref{app:regParam}) and to the force discretisation (see the rightmost column of table~\ref{tab:sph1e-6} for which the regularisation error is minimal); moreover it is advantageous to take \(h_f^{-1} h_q \) to be small, i.e.\ \(h_f>h_q\).

\subsection{A refinement heuristic}
For practical purposes we can therefore recommend the heuristic in table~\ref{tab:alg1}:
\begin{table}
  \centering{
    \parbox{0.8\textwidth}{
    \begin{enumerate}
      \item Choose \(\epsilon\) much smaller than the lengthscale of the problem geometry \(L\). Regularisation error will typically be linear in \(\epsilon\), so results which are required to be highly accurate will require a proportionately small value of \(\epsilon\).
      \item Generate the force discretisation -- initially this discretisation would be chosen relatively coarse.
      \item Generate the quadrature discretisation at least four times as fine as the force discretisation, i.e.\ \(h_q\) is no larger than \(h_f/4\).
      \item Assess convergence by halving \(h_f\), keeping \(h_q\) constant, and halving \(h_q\), keeping \(h_f\) constant.  Variations comparable to or smaller in magnitude than \(\epsilon\) are considered acceptable. Larger variations are unacceptable; halve \(h_f\) and \(h_q\) and repeat until convergence.
    \end{enumerate}
    }
  }
  \caption{Heuristic for calculating converged results.}\label{tab:alg1}
\end{table}
Discretisation convergence can then be assessed by (1) halving \(h_f\) while keeping \(h_q\) constant; (2) halving \(h_q\) while keeping \(h_f\) constant.

The heuristic in table~\ref{tab:alg1} can be applied to the rotating sphere problem as follows. We choose \(\epsilon=0.01\) as the regularisation parameter, and consider numerical errors comparable to \(1\%\) acceptable. Taking a relatively coarse force discretisation with \(h_f=0.5796\) and a finer quadrature discretisation of \(h_q=0.0416\) -- less than \(h_f/4\) -- we compute the total moment associated with the rigid body motion \(\bm{\Omega}=(1,0,0)\). We then assess convergence by halving each of \(h_f\) and \(h_q\). The results are shown in table~\ref{tab:sphAlgTest1}.

\begin{figure}
  \centering{
  \begin{tabular}{ll}
    (a) & (b) \\
    \includegraphics{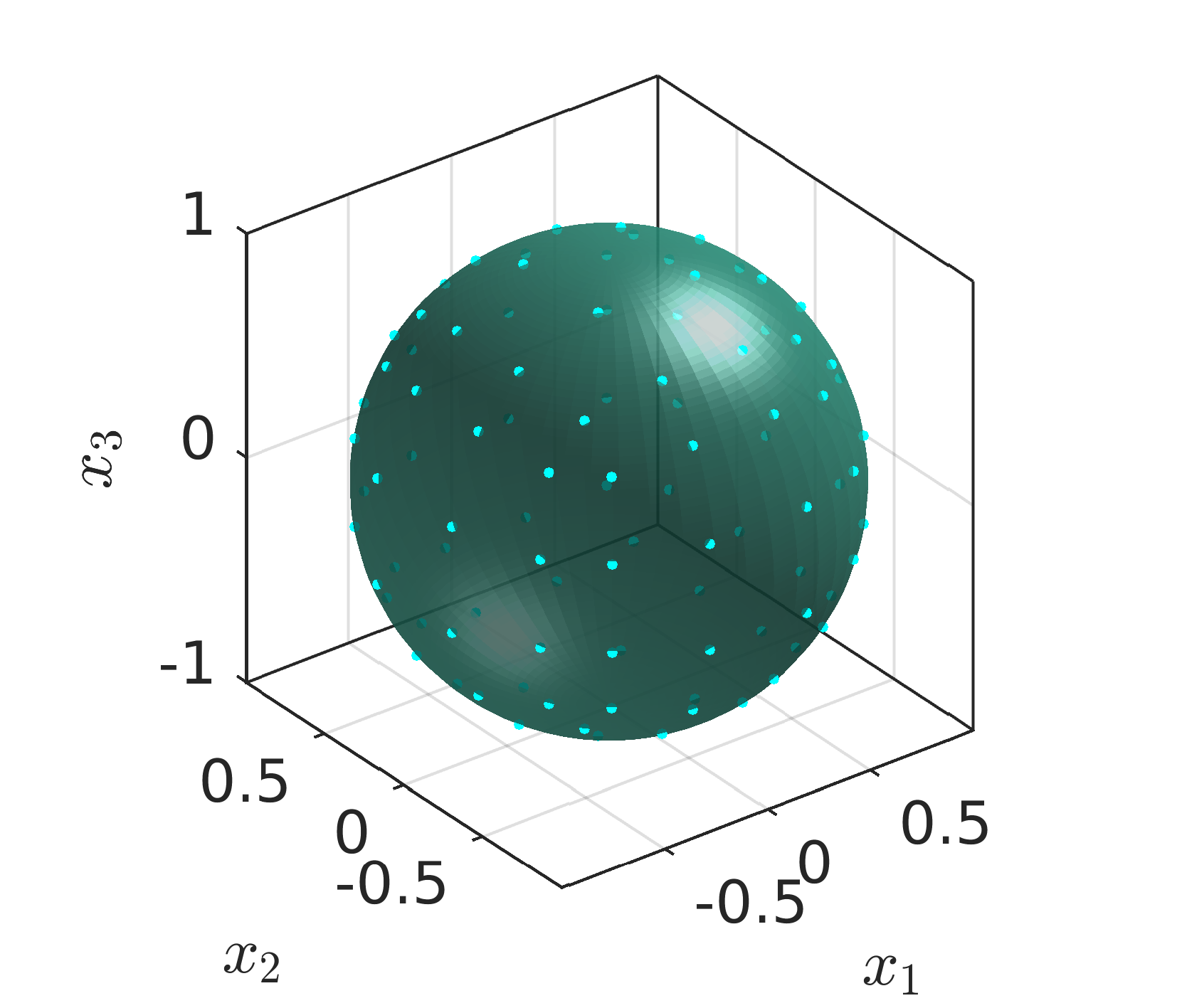} & \includegraphics{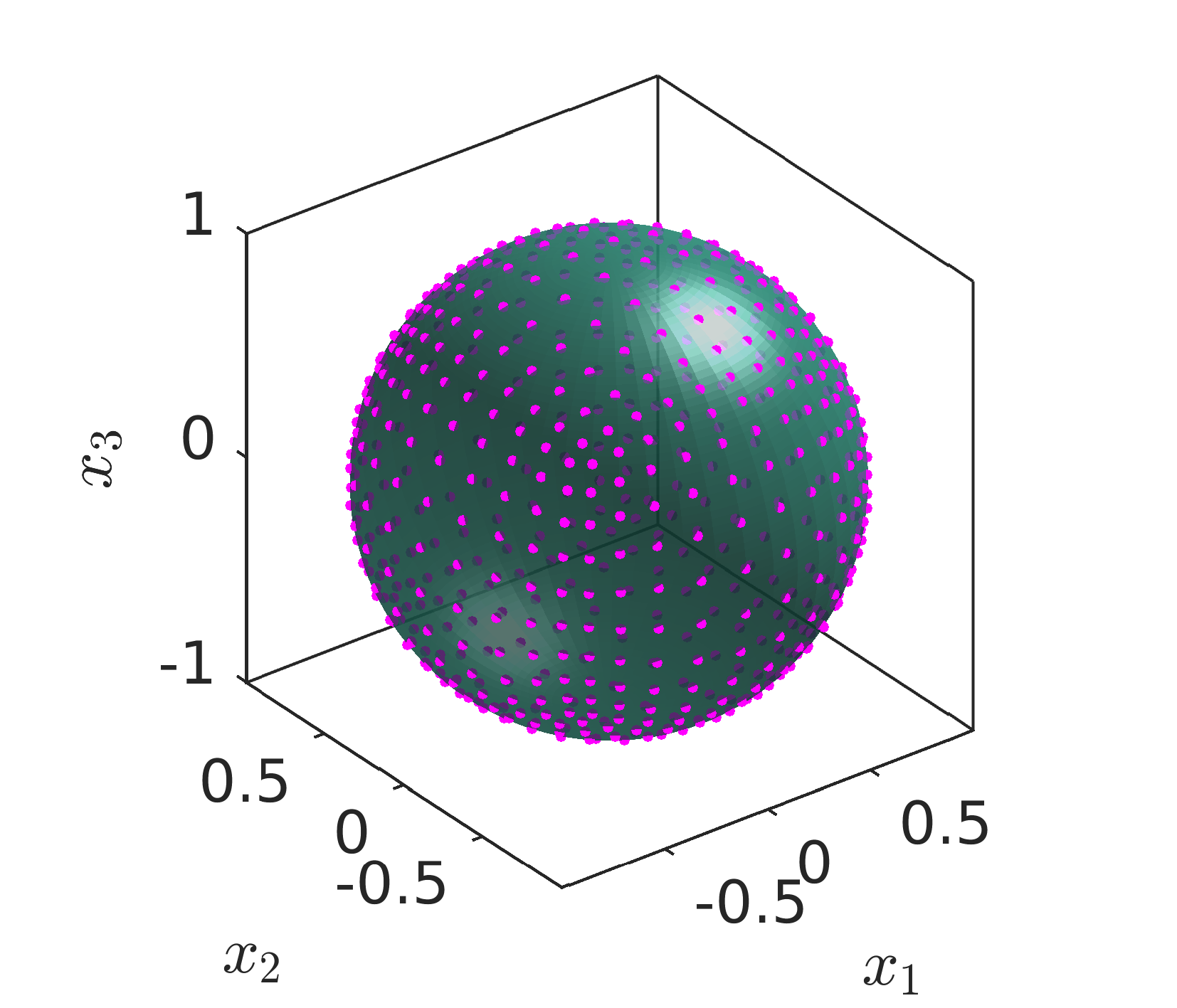}
  \end{tabular}
  }
  \caption{Visualisation of discretisations on the surface of a sphere: (a) force/collocation points with \(N=96\) (\(4\times 4\) subdivisions per face), (b) quadrature discretisation with \(Q=600\) (\(10\times 10\) subdivisions per face.) }  \label{fig:sphereDiscr}
\end{figure}

\begin{table}
  \centering{
  \begin{tabular}{l}
    (a)
    \\
    \begin{tabular}{cccc|ccccc}
              &     &           & \(Q   \)& 864& 3456& 13824& 55296& 221184\\
              &     &           & \(h_q \)& 0.1611& 0.0826& 0.0416& 0.0208& 0.0104\\
    \(N  \) & DOF & \(h_f\) &          \\[0.3em]  \hline
              &     &           &          \\[-0.8em]
    54 & 162 & 0.5796 & & 0.0147& 0.0052& 0.0002& 0.0006& 0.0012\\
    216 & 648 & 0.2942 & & 0.0166& 0.0079& 0.0038& 0.0022& 0.0020\\
    864 & 2592 & 0.1611 & & 0.1262& 0.0083& 0.0043& 0.0027& 0.0025\\
    3456 & 10368 & 0.0826 & & NaN& 0.0277& 0.0043& 0.0028& 0.0025\\
    \end{tabular}
    \vspace{0.5cm}
    \\
    (b)
    \\
    \begin{tabular}{cccc|ccccc}
              &     &           & \(Q   \)& 864& 3456& 13824& 55296& 221184\\
              &     &           & \(h_q \)& 0.1611& 0.0826& 0.0416& 0.0208& 0.0104\\
    \(N  \) & DOF & \(h_f\) &          \\[0.3em]  \hline
              &     &           &          \\[-0.8em]
    54 & 162 & 0.5796 & & 0.0300& 0.0086& 0.0019& 0.0036& 0.0047\\
    216 & 648 & 0.2942 & & 0.0378& 0.0182& 0.0095& 0.0063& 0.0058\\
    864 & 2592 & 0.1611 & & 0.2193& 0.0194& 0.0109& 0.0078& 0.0074\\
    3456 & 10368 & 0.0826 & & NaN& 0.0495& 0.0110& 0.0080& 0.0075\\
    \end{tabular}
  \end{tabular}
  }
  \caption{Relative error for the resistance problem of a unit sphere undergoing rigid body motion in Stokes flow in an infinite fluid; regularisation parameter \(\epsilon=0.01\). (a) Translation with velocity \(\bm{U}=(1,0,0)\). (b) Rotation with angular velocity \(\bm{\Omega}=(1,0,0)\).}\label{tab:sph001}
\end{table}

\begin{table}
  \centering{
    \begin{tabular}{ccc|ccccc}
              &     &    \(Q   \)& 864& 3456& 13824& 55296& 221184\\
    \(N  \) & DOF &           \\[0.3em]  \hline
              &     &           \\[-0.8em]
    54 & 162 &  & 132.807& 903.874& 415.837& 292.569& 260.382\\
    216 & 648 &  & 100.744& 242.112& 904.001& 3344.192& 1721.377\\
    864 & 2592 &  & 8.129& 218.433& 638.954& 4823.278& 15637.232\\
    3456 & 10368 &  & Inf & 38.309& 617.308& 3696.167& 1870680.776\\
    \end{tabular}

  }
    \caption{Condition number for the stokeslet matrix associated with the solution of the translation and rotation problems for a unit sphere in Stokes flow; regularisation parameter \(\epsilon=0.01\).}\label{tab:sphCond}
\end{table}

\begin{table}
  \centering{
    \begin{tabular}{cccc|cc}
              &     &           & \(Q   \)& 13824& 55296\\
              &     &           & \(h_q \)& 0.0416& 0.0208\\
    \(N  \) & DOF & \(h_f\) &          \\[0.3em]  \hline
              &     &           &          \\[-0.8em]
    54 & 162 & 0.5796 & & 25.0854& 25.0430\\
    216 & 648 & 0.2942 & & 25.3707& (25.2904) \\
    \end{tabular}

  }
    \caption{Results from applying heuristic~\ref{tab:alg1} to the problem of calculating the total moment on a unit sphere with unit angular velocity with \(\epsilon=0.01\). The result for \(h_f=0.5796\), \(h_q=0.0416\) is accurate to approximately 1\% relative error. The result shown in parentheses would not be calculated via this heuristic, but confirms the accuracy of the method.}\label{tab:sphAlgTest1}
\end{table}

\subsection{Rotational diffusion of a macromolecular structure}\label{metalloprotein}
\begin{figure}
\centering{
  \begin{tabular}{ll}
    (a) & (b) \\
    \includegraphics{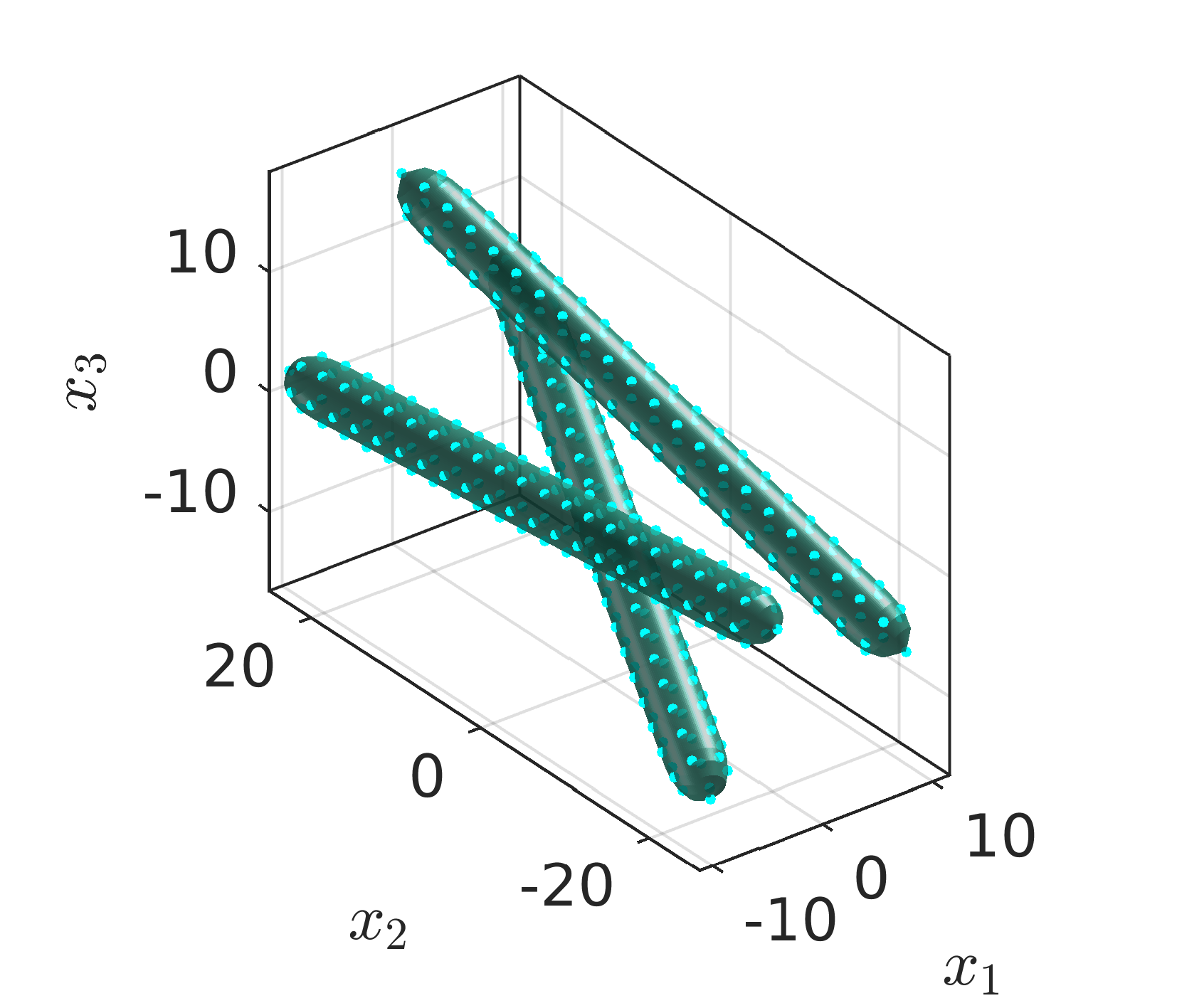}
    &
    \includegraphics{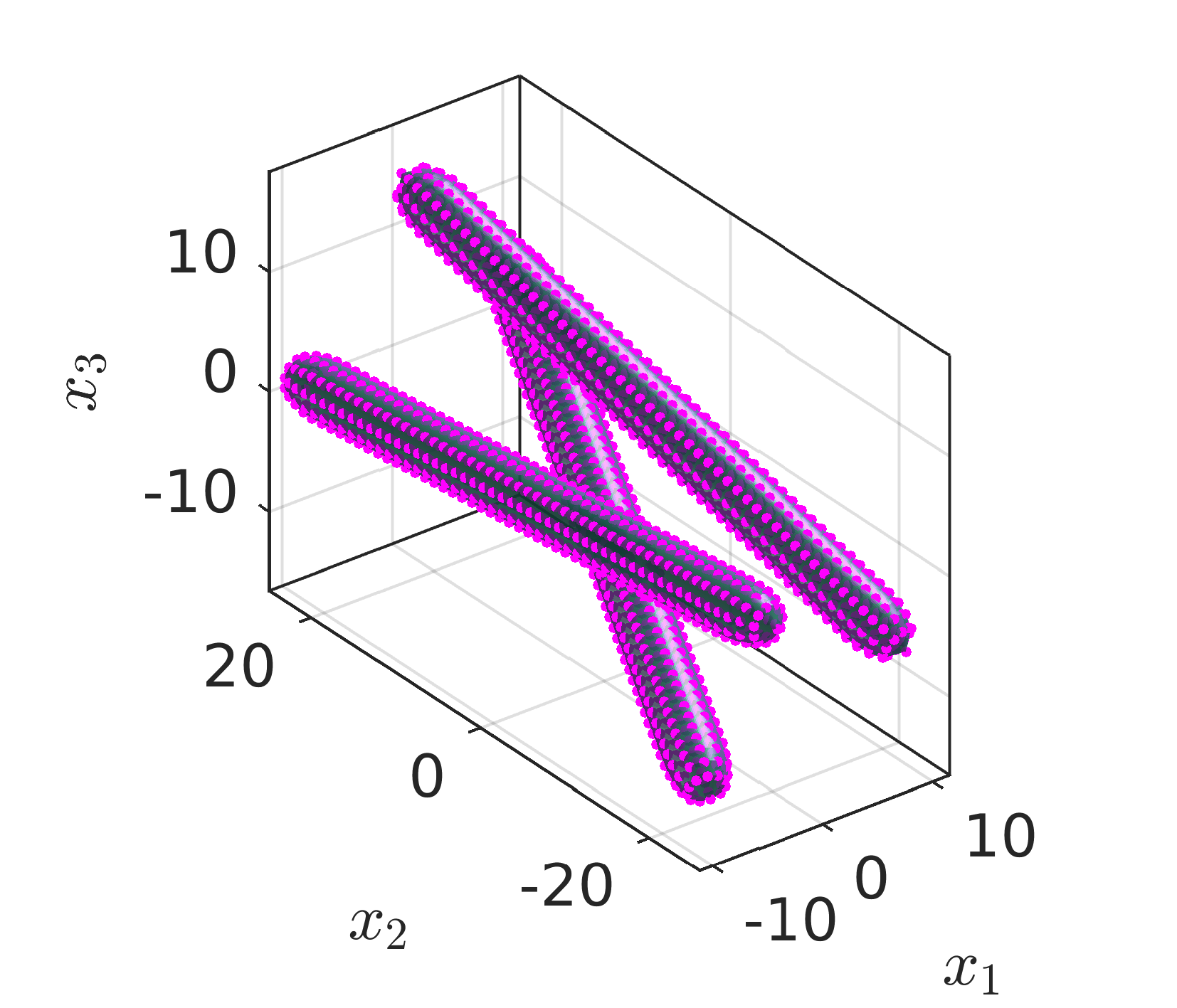}
  \end{tabular}
  }
\caption{Simplified representations of the macromolecular structure of interest, with (a) force discretisation (\(N=384\)) and (b) quadrature discretisation (\(Q=1689\)) shown.}\label{fig:metallopeptide}
\end{figure}

The technique will now be applied to a problem from bioinorganic chemistry: determining the rotational diffusion coefficient of a novel macromolecular structure. The scientific application of the calculations will be contained in a future colloborative publication. The structure can be modelled as three nanoscale rods with slightly different orientations, in close proximity, as shown in figure~\ref{fig:metallopeptide}, moving together as a single rigid body. The rods are discretised by subdividing equally in angle, and equally along the length of the rods; the angle and length spacings are chosen based on a target distance in both axial and azimuthal directions.

The grand resistance tensor \cite{pozrikidis2002} is defined as the \(6 \times 6\) matrix,
\begin{equation}
  \mathcal{R}=\begin{pmatrix} R_{FU} & R_{F\Omega} \\ R_{MU} & R_{M\Omega} \end{pmatrix}, \label{eq:grand}
\end{equation}
where \(R_{FU}\) is the force-velocity resistance matrix, \(R_{F\Omega}\) is the force-rotation coupling, \(R_{MU}\) is the moment-translation coupling and \(R_{M\Omega}\) is the moment-rotation resistance. This matrix relates the force \(\bm{F}\) and moment \(\bm{M}\) exerted by a rigid body on a viscous fluid to the body's translational velocity \(\bm{U}\) and angular velocity \(\bm{\Omega}\),
\begin{equation}
  \begin{pmatrix} \bm{F} \\ \bm{M} \end{pmatrix}
  =
  \mathcal{R} \begin{pmatrix} \bm{U} \\ \bm{\Omega} \end{pmatrix}.
\end{equation}
The individual components of the \(3\times 3\) matrices \(R_{\cdot\cdot}\) are calculated by solving the resistance problems \(\bm{U}=\bm{e}_j\) and \(\bm{\Omega}=\bm{e}_j\) in turn and calculating the force and moment in each case.

The diffusion tensor is given by \(\mathcal{D}=k_B T \mathcal{R}^{-1}\), where \(k_B\) is the Boltzmann constant and \(T\) is absolute temperature. The rotational part of the diffusion tensor is the lower right \(3\times 3\) block of \(\mathcal{D}\) \cite{wegener1981}, which we denote \(D_R\),
\begin{equation}
\mathcal{D} = \begin{pmatrix} D_T & D_C' \\ D_C & D_R \end{pmatrix}.
\end{equation}
The \(D_R\) block has no dependence on choice of origin \cite{wegener1981} (unlike the other blocks of \(\mathcal{D}\)); it has been verified numerically that moving the origin does not affect the calculation of \(D_R\).

It is convenient to report the smallest eigenvalue \(\lambda_1\) of \(D_R\), which corresponds to the smallest coefficient of rotational diffusion about each of the principal axes of rotation. The characteristic timescale of rotational diffusion is then given by \(\tau_1=1/(6\lambda_1)\). The results are given in table~\ref{tab:metallopeptide}(a).

Starting in the top left corner, applying our heuristic, repeating the process of dividing both \(h_f\) and \(h_q\) yields the values given on the main diagonal. The point to terminate the refinement process depends on the degree of accuracy required, and indeed if a relative error of less than 1\% is required, the process should be continued further. However for many biophysical applications, the level of modelling error (for example, approximating the structure by three straight rigid rods, assuming rigidity) does not warrant extremely precise numerical calculations. Fixing \(h_q\) and examining the first three columns, the nearest-neighbour method with coarser force discretisations \(h_f\approx 1.5h_q\)--\(2.4h_q\) out-performs the Nystr\"{o}m method (\(h_f=h_q\)) for both accuracy and efficiency (table~\ref{tab:metallopeptide}(b)).

\begin{table}
  \centering{
  \begin{tabular}{c}
   \begin{tabular}{l}
    (a)
    \\
    \begin{tabular}{cccc|ccc}
              &     &           & \(Q   \)& 3270& 12678& 49926\\
              &     &           & \(h_q \)& 1.5544& 1.0543& 0.7288\\
    \(N  \) & DOF & \(h_f\) &          \\[0.3em]  \hline
              &     &           &          \\[-0.8em]
    246 & 738 & 3.7165 & & 5.9021& 6.0156& 5.9831\\
    870 & 2610 & 2.3598 & & 5.7480& 5.9255& 5.9160\\
    3270 & 9810 & 1.5544 & & 5.6406& 5.8281& 5.8916\\
    \end{tabular}
   \end{tabular}
   \vspace{0.5cm}
    \\
   \begin{tabular}{l}
    (b)
    \\
    \begin{tabular}{ccc|ccc}
              &     &    \(Q   \)& 3270& 12678& 49926\\
    \(N  \) & DOF &           \\[0.3em]  \hline
              &     &           \\[-0.8em]
    246 & 738 &  & 2.542& 11.814& 99.667\\
    870 & 2610 &  & 10.526& 34.312& 173.740\\
    3270 & 9810 &  & 125.025& 316.446& 742.062\\
    \end{tabular}
   \end{tabular}
  \end{tabular}
  }
  \caption{Calculation of the rotational diffusion timescale \(\tau_1\) for the macromolecular model shown in figure~\ref{fig:metallopeptide}. The regularisation parameter \(\epsilon\) is taken as \(0.01L\) where \(L\) is the approximate half-length of the peptide, 25 \AA. The absolute temperature \(T=310\)~K and dynamic viscosity \(\mu=10^{-3}\)~Pa.s. Results with \(h_f=h_q\) (second sub-diagonal) relate to the classic Nystr\"{o}m discretisation, results with \(h_f<h_q\) relate to `nearest-neighbour' discretisations. (a) Rotational diffusion timescale \(\tau_1\) in nanoseconds for each discretisation tested; discretisation parameters \(h_f\) and \(h_q\) are given in \AA. (b) Computational timings (in seconds; notebook specification given in appendix~\ref{app:sphTimings}).}\label{tab:metallopeptide}
\end{table}

\section{Conclusions}
We have presented a simple-to-implement modification of the standard discretisation of the method of regularized stokeslets for modelling particle dynamics at zero Reynolds number. The modification is based on the use of two discretisations, one for the unknown surface force per unit area and one for the stokeslet quadrature, combined with nearest-neighbour discretisation of the force distribution. Practically, the method can be implemented by assembling a nearest-neighbour operator matrix, which can be achieved with a few lines of \matlab/GNU Octave code. Numerical experiments on the resistance problem of a sphere undergoing rigid body motion, and the calculation of the rotational diffusion timescale of a macromolecular structure provide evidence that the method enables more accurate results to be obtained at lower computational cost than the standard implementation, despite not being substantially more complicated to implement. Our initial error estimate \(O(\epsilon)+O(\epsilon^{-2} h_f^2 h_q)+O(h_f^{-1} h_q) + O(h_f)\) provides insight into the independent effects of \(h_f\) and \(h_q\) on the numerical error, and the potential advantage of taking \(h_f>h_q\), provided that \(h_f\) is not too large. Numerical results did not however reflect the sensitivity to \(\epsilon\) suggested by this estimate -- further investigation of this phenomenon, and possible sharpening of the estimate, may be topics for future work.

The standard Nystr\"{o}m discretisation uses, in our framework, the same discretisation for the force and the quadrature. The present approach shows that this choice much less reliably produces accurate results than if the quadrature discretisation is kept the same but the force discretisation is made twice as coarse. Making the force discretisation twice as coarse means that the number of degrees of freedom is at least halved (more typically reduced by a factor of four). The matrix assembly cost is therefore reduced by a factor of at least four, and the linear solver cost is reduced by a factor of at least eight (for a direct solver). Our practical results suggest a cost reduction of over 10 times may be typical. This reduction in cost means that more complex problems can be solved with a given computational resource -- a useful facet, particularly within biological and biophysical fluid dynamics. The code implementation described in this paper makes use primarily of basic linear algebra operations rather than serial for-loops, and therefore can be accelerated through built-in software and hardware parallelisation of these operations. It will be of interest to explore how the algorithm scales on multicore or GPU hardware.

The nearest-neighbour approach still has limitations, particularly if compared with boundary element methods -- which may involve higher order force discretisation and adaptive quadrature, and accelerations such as the fast multipole method. However, the nearest-neighbour approach is very simple to implement, requiring only a small modification of the standard regularized stokeslet approach, and not requiring true mesh generation. It may be valuable to explore further whether adaptivity or fast multipole implementations can be introduced without excessively complicating the algorithm. Finally, we do not yet have theoretical results which definitively prove the improved efficiency and accuracy of the method. Nevertheless, for practical purposes, carrying out a sequence of discretisations with \(h_f\approx 4h_q\) alongside a sequence with \(h_f\approx 2h_q\) will establish convergence empirically.

The nearest-neighbour discretisation of the regularized stokeslet method is more efficient and accurate than the standard implementation, with minimal additional complexity. It may therefore enable researchers in biological and biophysical fluid dynamics to solve significantly more challenging open problems, for example involving many swimming cells, ciliated cavities, and/or suspended macromolecules. The task of explaining the properties of the method, which we are only able to explain heuristically at present, may stimulate theoretical work. Finally, the technique may also open the way for future algorithmic developments which possess the efficiency and accuracy of boundary element methods but retain its useful properties of meshlessness and simplicity.

\section*{Acknowledgments}
This research was supported by Engineering and Physical Sciences Research Council grants EP/K007637/1 and EP/N021096/1. The author acknowledges Drs Anna Peacock and Sarah Newton (University of Birmingham) for suggesting the macromolecular diffusion problem in section~\ref{metalloprotein}, valuable discussions about regularized stokeslet methods and diffusion tensor calculations with Dr Rudi Schuech (University of Lincoln), and valuable comments from two anonymous referees).

\begin{appendices}

\section{\matlab/GNU Octave implementation}\label{app:code}
The essentials of the \matlab/GNU Octave implementation are given below, in particular some more subtle aspects such as the assembly of the nearest-neighbour matrix, avoidance of extensive for-loops, and use of `blocking' to avoid memory overrun.

\subsection{Regularized Stokeslet matrix}\label{app:regsto}

Taking advantage of the vectorisation capabilities of the \matlab language and the Kronecker product operator,
\begin{lstlisting}
function S=RegStokeslet(x,X,ep)
	% x is a vector of field points:  3*M
	% X is a vector of source points: 3*Q
	% ep is regularisation parameter
        % outputs an array of regularized stokeslets between field
        %        and source points
        % blocks are [Sxx, Sxy, Sxz; Syx, Syy, Syz; Szx, Szy, Szz]
        %        where Sxx is M by Q etc.
	x=x(:);
	X=X(:);
	M=length(x)/3;
	Q=length(X)/3;
	r1=      x(1:M)*ones(1,Q)-ones(M,1)*      X(1:Q)';
	r2=  x(M+1:2*M)*ones(1,Q)-ones(M,1)*  X(Q+1:2*Q)';
	r3=x(2*M+1:3*M)*ones(1,Q)-ones(M,1)*X(2*Q+1:3*Q)';
	rsq=r1.^2+r2.^2+r3.^2;
	irep3=1./(sqrt((rsq+ep^2)).^3);
	isotropic=kron(eye(3),(rsq+2.0*ep^2).*ireps3);
	dyadic=[r1.*r1 r1.*r2 r1.*r3; r2.*r1 r2.*r2 r2.*r3; ...
                 r3.*r1 r3.*r2 r3.*r3].*kron(ones(3,3),irep3);
	S=(1.0/(8.0*pi))*(isotropic+dyadic);
\end{lstlisting}

\subsection{Nearest-neighbour matrix}\label{app:nnmat}
The nearest-neighbour operator \(\nu\) may be discretised with the matrix \verb@NN@ produced by the following code,
\begin{lstlisting}
function NClosest=NearestNeighbourMatrix(X,x,varargin)
        % Vectors should be supplied with all x1 coordinates listed
        % first then all x2 coordinates, then all x3 coordinates.
        % if varargin is nonempty, then it should contain
        % blockSize
        Q=length(X)/3;N=length(x)/3;
        if ~isempty(varargin)
            blockSize=varargin{1};
            blockNodes=floor(blockSize*2^27/(9*N));
        else
            blockNodes=Q;
        end
        xQ1=X(1:Q);
        xQ2=X(Q+1:2*Q);
        xQ3=X(2*Q+1:3*Q);
        xT1=x(1:N);
        xT2=x(N+1:2*N);
        xT3=x(2*N+1:3*N);
        nMin=zeros(Q,1);
        for iMin=1:blockNodes:Q
            iMax=min(iMin+blockNodes-1,Q);
            blockCurr=iMax-iMin+1;
            X1=xQ1(iMin:iMax)*ones(1,N)-ones(blockCurr,1)*xT1';
            X2=xQ2(iMin:iMax)*ones(1,N)-ones(blockCurr,1)*xT2';
            X3=xQ3(iMin:iMax)*ones(1,N)-ones(blockCurr,1)*xT3';
            distsq=X1.^2+X2.^2+X3.^2;
            [~,nMin(iMin:iMax)]=min(distsq,[],2);
        end
        NClosest=sparse(Q,N); % creates sparse all-zero matrix
        NClosest([1:Q]'+Q*(nMin-1))=1;
        NClosest=kron(speye(3),NClosest);
\end{lstlisting}

The above takes advantage of the speed of predominantly vector operations, whilst not exceeding the memory requirements of the system. The optional third argument, \verb@blockSize@ is a measurement in GB of the memory to be allocated to the regularized stokeslet matrix so that
\begin{lstlisting}
        blockNodes=floor(blockSize*2^27/(9*N));
\end{lstlisting}
gives the number of columns (corresponding to a subset of the force points) which can be dealt with simultaneously. For example, \verb@blockSize=0.2@ would be suitable for any modern hardware, and has been tested on a Raspberry Pi Model B.
The matrix \verb@NClosest@, which corresponds to \(\nu[q,n]\) is sparse and so will not produce a memory overflow. The final line involving the Kronecker product operation is required because the nearest-neighbour operator must be copied into three blocks, acting on the \(f_1\), \(f_2\) and \(f_3\) components in turn.

\subsection{Resistance problem}\label{app:resprob}
The `left hand side' matrix \(\mathsf{A}\) for the discrete resistance problem~\eqref{eq:nnres2} can then be assembled as,
\begin{lstlisting}
        A = RegStokeslet(x,X,ep)*NearestNeighbourMatrix(X,x);
\end{lstlisting}
The regularized stokeslet matrix may be too large to fit in memory, particularly if \(Q\) is very large, as may be the case for problems possessing complex geometry. In this case, the problem may be assembled `block-by-block' as follows,
\begin{lstlisting}
        NN=NearestNeighbourMatrix(X,x,blockSize);
        A=zeros(3*M,3*N);
        for iMin=1:blockNodes:Q
           iMax=min(iMin+blockNodes-1,Q);
           iRange=[iMin:iMax Q+iMin:Q+iMax 2*Q+iMin:2*Q+iMax];
           A=A+RegStokeslet(x,X(iRange),ep)*NN(iRange,:);
        end
\end{lstlisting}
As in the function \verb@NearestNeighbour@, blocking is used to prevent overrun. In all calculations in the present report, the linear system was solved with the `backslash' operator, i.e.\ \verb@f=A\b@

\subsection{Discretisation size calculation}\label{app:calcdiscr_h}
The discretisation size parameters \(h_f\) and \(h_q\) are calculated using the following function,
\begin{lstlisting}
function [h,hMin,nMin,distsq] = CalcDiscr_h(x)
% CalcDiscr_h This function calculates the maximum over all
%       points in a discretisation x of the distance to
%       the nearest-neighbour point
        N=length(x)/3;
        X1=x(1:N)*ones(1,N)-ones(N,1)*x(1:N)';
        X2=x(N+1:2*N)*ones(1,N)-ones(N,1)*x(N+1:2*N)';
        X3=x(2*N+1:3*N)*ones(1,N)-ones(N,1)*x(2*N+1:3*N)';
        distsq=X1.^2+X2.^2+X3.^2+100*eye(N);
        [hMin,nMin]=min(distsq,[],2);
        h=sqrt(max(hMin));
\end{lstlisting}

\section{Effect of the regularization parameter for the rigid sphere test problem}\label{app:regParam}
To assess the sensitivity of the method to regularisation parameter, we present test results with \(\epsilon=0.02\) and \(\epsilon=0.005\) in tables~\ref{tab:sph002} and \ref{tab:sph0005} respectively. When \(h_f\) and \(h_q\) are taken equal, the results are highly sensitive to the value of \(\epsilon\), however provided \(h_q\) is taken no larger than \(0.25h_f\), the error is relatively insensitive. As \(\epsilon\) is reduced, the finite regularisation error (evident in the rightmost entries in the tables) is reduced to below 1\%, however convergence to the smaller error with \(h_q\) is slower. Perhaps surprisingly, it does not appear necessary to choose \(h_q\) dependent on \(\epsilon\), at least within the range of values explored. There also does not appear to be any clear advantage to taking \(\epsilon=0.02\) as opposed to \(\epsilon=0.005\), begging the question of how small \(\epsilon\) can be taken. While \(\epsilon=0\) is equivalent to non-regularized stokeslets (and hence singular matrix entries whenever the collocation and quadrature points coincide), taking a very small but finite value of \(\epsilon=10^{-6}\) yields the results of table~\ref{tab:sph1e-6}, which are typically at least as accurate as the results with larger value of \(\epsilon\), provided that \(h_f\geqslant 2 h_q\).

\begin{table}
  \centering{
  \begin{tabular}{l}
    (a)
    \\
    \begin{tabular}{cccc|ccccc}
              &     &           & \(Q   \)& 864& 3456& 13824& 55296& 221184\\
              &     &           & \(h_q \)& 0.1611& 0.0826& 0.0416& 0.0208& 0.0104\\
    \(N  \) & DOF & \(h_f\) &          \\[0.3em]  \hline
              &     &           &          \\[-0.8em]
    54 & 162 & 0.5796 & & 0.0149& 0.0056& 0.0016& 0.0017& 0.0014\\
    216 & 648 & 0.2942 & & 0.0167& 0.0083& 0.0052& 0.0046& 0.0046\\
    864 & 2592 & 0.1611 & & 0.0522& 0.0087& 0.0056& 0.0051& 0.0051\\
    3456 & 10368 & 0.0826 & & NaN& 0.0046& 0.0057& 0.0052& 0.0051\\
    \end{tabular}
    \vspace{0.5cm}
    \\
    (b)
    \\
    \begin{tabular}{cccc|ccccc}
              &     &           & \(Q   \)& 864& 3456& 13824& 55296& 221184\\
              &     &           & \(h_q \)& 0.1611& 0.0826& 0.0416& 0.0208& 0.0104\\
    \(N  \) & DOF & \(h_f\) &          \\[0.3em]  \hline
              &     &           &          \\[-0.8em]
    54 & 162 & 0.5796 & & 0.0313& 0.0114& 0.0034& 0.0036& 0.0030\\
    216 & 648 & 0.2942 & & 0.0391& 0.0210& 0.0147& 0.0136& 0.0135\\
    864 & 2592 & 0.1611 & & 0.0917& 0.0222& 0.0161& 0.0151& 0.0150\\
    3456 & 10368 & 0.0826 & & NaN& 0.0032& 0.0162& 0.0152& 0.0152\\
    \end{tabular}
  \end{tabular}
  }
  \caption{Relative error for the resistance problem of a unit sphere undergoing rigid body motion in Stokes flow in an infinite fluid; regularisation parameter \(\epsilon=0.02\). (a) Translation with velocity \(\bm{U}=(1,0,0)\). (b) Rotation with angular velocity \(\bm{\Omega}=(1,0,0)\). `NaN' denotes `not-a-number', and indicates a singular linear system.}\label{tab:sph002}
\end{table}

\begin{table}
  \centering{
  \begin{tabular}{l}
    (a)
    \\
    \begin{tabular}{cccc|ccccc}
              &     &           & \(Q   \)& 864& 3456& 13824& 55296& 221184\\
              &     &           & \(h_q \)& 0.1611& 0.0826& 0.0416& 0.0208& 0.0104\\
    \(N  \) & DOF & \(h_f\) &          \\[0.3em]  \hline
              &     &           &          \\[-0.8em]
    54 & 162 & 0.5796 & & 0.0147& 0.0051& 0.0000& 0.0013& 0.0023\\
    216 & 648 & 0.2942 & & 0.0165& 0.0079& 0.0036& 0.0016& 0.0008\\
    864 & 2592 & 0.1611 & & 0.2424& 0.0082& 0.0041& 0.0021& 0.0013\\
    3456 & 10368 & 0.0826 & & NaN& 0.0691& 0.0041& 0.0021& 0.0014\\
    \end{tabular}
    \vspace{0.5cm}
    \\
    (b)
    \\
    \begin{tabular}{cccc|ccccc}
              &     &           & \(Q   \)& 864& 3456& 13824& 55296& 221184\\
              &     &           & \(h_q \)& 0.1611& 0.0826& 0.0416& 0.0208& 0.0104\\
    \(N  \) & DOF & \(h_f\) &          \\[0.3em]  \hline
              &     &           &          \\[-0.8em]
    54 & 162 & 0.5796 & & 0.0297& 0.0079& 0.0033& 0.0062& 0.0082\\
    216 & 648 & 0.2942 & & 0.0375& 0.0176& 0.0081& 0.0037& 0.0021\\
    864 & 2592 & 0.1611 & & 0.3876& 0.0188& 0.0095& 0.0053& 0.0038\\
    3456 & 10368 & 0.0826 & & NaN& 0.1262& 0.0096& 0.0054& 0.0040\\
    \end{tabular}
  \end{tabular}
  }
  \caption{Relative error for the resistance problem of a unit sphere undergoing rigid body motion in Stokes flow in an infinite fluid; regularisation parameter \(\epsilon=0.005\). (a) Translation with velocity \(\bm{U}=(1,0,0)\). (b) Rotation with angular velocity \(\bm{\Omega}=(1,0,0)\).}\label{tab:sph0005}
\end{table}

\begin{table}
  \centering{
  \begin{tabular}{l}
    (a)
    \\
    \begin{tabular}{cccc|ccccc}
              &     &           & \(Q   \)& 864& 3456& 13824& 55296& 221184\\
              &     &           & \(h_q \)& 0.1611& 0.0826& 0.0416& 0.0208& 0.0104\\
    \(N  \) & DOF & \(h_f\) &          \\[0.3em]  \hline
              &     &           &          \\[-0.8em]
    54 & 162 & 0.5796 & & 0.0147& 0.0051& 0.0000& 0.0014& 0.0027\\
    216 & 648 & 0.2942 & & 0.0165& 0.0079& 0.0036& 0.0015& 0.0004\\
    864 & 2592 & 0.1611 & & 0.9994& 0.0082& 0.0040& 0.0020& 0.0009\\
    3456 & 10368 & 0.0826 & & NaN& 0.9977& 0.0041& 0.0020& 0.0010\\
    \end{tabular}
    \vspace{0.5cm}
    \\
    (b)
    \\
    \begin{tabular}{cccc|ccccc}
              &     &           & \(Q   \)& 864& 3456& 13824& 55296& 221184\\
              &     &           & \(h_q \)& 0.1611& 0.0826& 0.0416& 0.0208& 0.0104\\
    \(N  \) & DOF & \(h_f\) &          \\[0.3em]  \hline
              &     &           &          \\[-0.8em]
    54 & 162 & 0.5796 & & 0.0296& 0.0077& 0.0037& 0.0071& 0.0100\\
    216 & 648 & 0.2942 & & 0.0374& 0.0174& 0.0077& 0.0028& 0.0004\\
    864 & 2592 & 0.1611 & & 0.9997& 0.0186& 0.0091& 0.0044& 0.0021\\
    3456 & 10368 & 0.0826 & & NaN& 0.9988& 0.0092& 0.0046& 0.0023\\
    \end{tabular}
  \end{tabular}
  }
  \caption{Relative error for the resistance problem of a unit sphere undergoing rigid body motion in Stokes flow in an infinite fluid; regularisation parameter \(\epsilon=10^{-6}\). (a) Translation with velocity \(\bm{U}=(1,0,0)\). (b) Rotation with angular velocity \(\bm{\Omega}=(1,0,0)\).}\label{tab:sph1e-6}
\end{table}

\section{Timing results for the rigid sphere test problem}\label{app:sphTimings}
Typical timing results (in seconds) for the solution of the translation and rotation resistance problems (with \(\epsilon=0.01\)) computed on a modest notebook computer (2011 Lenovo Thinkpad X220; Intel(R) Core(TM) i5-2520M CPU @ 2.50GHz; 8GB DDR3 RAM) are given in table~\ref{tab:sphTimings}.

\begin{table}
  \centering{
    \begin{tabular}{ccc|ccccc}
              &     &    \(Q   \)& 864& 3456& 13824& 55296& 221184\\
    \(N  \) & DOF &           \\[0.3em]  \hline
              &     &           \\[-0.8em]
    54 & 162 &  & 0.052& 0.455& 5.288& 81.071& 1319.765\\
    216 & 648 &  & 0.287& 0.882& 6.471& 84.248& 1318.965\\
    864 & 2592 &  & 5.773& 8.726& 15.953& 105.991& 1399.280\\
    3456 & 10368 &  & 352.833& 320.650& 373.632& 535.916& 2160.078\\
    \end{tabular}
    }
\caption{Timing results for the calculation of the translation and rotation resistance problems for the unit sphere with \(\epsilon=0.01\).}\label{tab:sphTimings}
\end{table}

\section{Testing the method on a prolate spheroid}\label{app:prolate}
To explore further whether the efficiency of the choice \(h_f\approx 2 h_q\) is problem-dependent, we may assess the performance of the nearest-neighbour method in calculating the grand resistance tensor \(\mathcal{R}\) (defined in equation~\eqref{eq:grand}) of a rigid prolate spheroid, which has a well-known analytical solution \cite[p.\ 64]{kim1991}. Taking a prolate spheroid with long semi-axis \(a=5\) and short semi-axis \(c=1\), the relative error in \(\mathcal{R}\) in the \(\|\cdot\|_2\) norm is given in table~\ref{tab:prolate}. The results are not likely to be optimal as the discretisation has been created by simply deforming the sphere discretisation depicted in figure~\ref{fig:sphereDiscr} without any attempt to space the points uniformly in the directions of the long and short semi-axes.

\begin{table}
  \centering{
  \begin{tabular}{cccc|ccccc}
    &     &           & \(Q   \)& 864& 3456& 13824& 55296& 221184\\
    &     &           & \(h_q \)& 0.2171& 0.1117& 0.0554& 0.0278& 0.0139\\
    \(N  \) & DOF & \(h_f\) &          \\[0.3em]  \hline
    &     &           &          \\[-0.8em]
    54 & 162 & 1.0064 & & 0.0596& 0.0221& 0.0096& 0.0045& 0.0052\\
    216 & 648 & 0.4305 & & 0.0459& 0.0180& 0.0079& 0.0036& 0.0021\\
    864 & 2592 & 0.2171 & & 0.3675& 0.0207& 0.0097& 0.0049& 0.0034\\
    3456 & 10368 & 0.1117 & & NaN& 0.1171& 0.0099& 0.0052& 0.0036\\
  \end{tabular}
  }
  \caption{Test results for the grand resistance tensor of rigid body motion of a prolate spheroid with long semi-axis \(5\), short semi-axis \(1\); regularisation parameter \(\epsilon=0.01\).}\label{tab:prolate}
\end{table}

\end{appendices}

\end{document}